\documentstyle[aps,multicol,epsf]{revtex}

\begin{document}
\draft
\title{Evolution of a sandpile in a thick flow regime}

\author{S.N. Dorogovtsev$^{1, 2, \ast}$ and J.F.F. Mendes$^{1,\dagger}$}

\address{
$^{1}$ Departamento de F\'\i sica and Centro de F\'\i sica do Porto, Faculdade 
de Ci\^encias, 
Universidade do Porto\\
Rua do Campo Alegre 687, 4169-007 Porto, Portugal\\
$^{2}$ A.F. Ioffe Physico-Technical Institute, 194021 St. Petersburg, Russia 
}

\maketitle

\begin{abstract}
We solve a one-dimensional sandpile problem analytically in a thick flow regime 
when the pile evolution may be described by a set of linear equations. We 
demonstrate that, if an income flow is constant, a space periodicity takes place 
while the sandpile evolves even for a pile of only one type of particles. Hence, 
grains are piling layer by layer. The thickness of the layers is proportional to 
the input flow of particles $r_0$ and coincides with the thickness of stratified 
layers in a two-component sandpile problem which were observed recently. We find 
that the surface angle $\theta$ of the pile reaches its final critical value 
($\theta_f$) only at long times after a complicated relaxation process. The 
deviation  ($\theta_f  -  \theta $) behaves asymptotically as $(t/r_{0})^{-
1/2}$. It appears that the pile evolution depends on initial conditions. We 
consider two cases: (i) grains are absent at the initial moment, and (ii) there 
is already a pile with a critical slope initially. Although at long times the 
behavior appears to be similar in both cases, some differences are observed for  
the different initial conditions are observed. We show that the periodicity 
disappears if the input flow increases with time.
\end{abstract}


\begin{multicols}{2}

\narrowtext

\section{Introduction}

Granular flows have attracted increasing interest over the last years, and 
constitute now a very active research field with apparent technological applications \cite{beh,kadan,ed1,her1,g1,duran,jaeger,genn,frette}. 
Reasons for such great attractiveness of granular media or of the so called 
soft matter are clear: they present not only unusual 
properties which let them look like solids in some cases 
and liquids in other ones, but they also display new phenomena unknown both to 
solids and liquids. Such a new and intriguing phenomenon is recently observed 
spontaneous stratification of granular mixtures 
\cite{makse97n,makse97prl,makse97pre,makse981,ciz,makse983,karol}. When a 
granular mixture is poured into a Hele-Shaw cell (i.e. in a quasi-two dimensional 
silo) under some conditions, layers of different grains appear to be arranged 
periodically.

Several studies were carried out to explain the striking stratification 
\cite{makse97n,makse97prl,makse97pre,makse981,ciz,makse983,karol} although the 
problem seems to be very complicated for an analytical treatment. In our 
communication \cite{dm} we put forward the following question: is it possible to 
find some precursor of stratification in a much simpler situation when only one 
type of grains is poured? We shall present a full answer below for one of the limit 
regimes of sandpile spilling using an approach most convenient for analytical 
consideration. 

Crucial progress in understanding of granular flow nature was made some years 
ago when Bouchaud, Cates, Prakash, and Edwards \cite{bouch} (see also the papers 
of Mehta and Barker \cite{mehta} and of Mehta, Needs, and Dattagupta 
\cite{mehta1}) introduced a clear phenomenological theory that describes a 
surface flow of granular materials. The main idea of the approach is the 
following. The grains are divided to two parts -- static grains and rolling 
ones. The conversion of static grains to rolling ones occurs depending on the 
relation between the local slope of a pile and the repose angle of the material. 
In addition, the continuity equation for the total amount of grains holds. Thus, the 
problem can be reduced to two coupled partial differential equations for the local 
amounts of static and rolling grains. One should note that similar ideas were 
also applied to a self-organized criticality problem 
\cite{vesp97,dick98,vesp98}. 

Nevertheless, even this simple approach is still complicated because of 
nonlinearity of the equations.
Recently, Boutreux, Rapha\"{e}l and P.-G. de Gennes \cite{bout1} proposed a 
phenomenological description of some special case of granular flows  -- a so 
called thick flow regime -- that provides a unique possibility for an analytical 
treatment: in this case the coupled equations are linear and one of them is 
completely decoupled from the other one. In papers \cite{bout1,bout2} this 
approach was used to study some granular flow configurations. Relaxation of sand 
from several most simple states was considered, generalization to many-component 
flows were proposed 
\cite{bout01,bout02,bout03}, and new improvements of the approach were made 
\cite{aradian} to make it closer to reality, though the main classical sandpile 
problem still remained unsolved even for the simplest version of such a 
description.   

We shall use the proposed phenomenological equations to answer positively the 
above stated question and to describe a total evolution of the sandpile, that 
will turn to be surprisingly complex. We will show that the slope of the pile 
approaches its critical value only at long times after a complicated 
discontinuous relaxation process \cite{dm}. In fact, we shall reconsider a 
classical sandpile problem using, maybe, the simplest possible idea without 
appealing to more refined approaches like, for instance, self-organized 
criticality \cite{bak,dhar,frette}.

To start with, let us write out the phenomenological equations we shall use. 
Boutreux, Rapha\"{e}l and P.-G. de Gennes  \cite{bout1} describe 
phenomenologically the one-dimensional thick granular flows (the flow thickness 
supposed to be much higher than the grain size) by the following equations:

\begin{eqnarray}
\label{e1}
\frac{\partial r}{\partial t} - v \; \frac{\partial r}{\partial x}   & = &  
v_{u} \left( \frac{\partial h}{\partial x} - \theta_{f} \right) , \nonumber \\
\frac{\partial h}{\partial t} &  =  & - v_{u} \left( \frac{\partial h}{\partial 
x} - \theta_{f} \right) .
\end{eqnarray}
Here $h(x,t)$ is a profile of the static part of the material; $r(x,t)$ is the 
width of a moving granular layer; we assume that the flow is from right to left 
and all rolling grains are supposed to move with an equal velocity $v$. $r(x,t)$ and 
$h(x,t)$ describe completely the evolution of a pile.
The sum of Eqs. (\ref{e1}), $\partial(r+h)/\partial t-v\partial r/\partial x = 
0$, has the form of the continuity equation for grains, $v r(x,t)$ is a local 
flow. The right hand parts describe the conversion of the static grains to 
rolling ones and vice versa depending on a relation between a local slope 
$\partial h/\partial x$ and $\theta_f$. The meaning of $v_u$ is the velocity of 
the uphill fronts as we shall see later. Usually $v_{u} > v$ \cite{ciz,bout1}. 
$\theta_f$ is a critical angle, or a so called repose angle, that is the angle 
to which the sandpile will evolve. The deviations of the local slope from 
$\theta_f$ are supposed to be small. Eq. (\ref{e1}) is valid for $r(x,t)>0$. If 
$r(x,t)=0$, then the right hand parts of Eqs. (\ref{e1}) are supposed to be equal to zero. 
Therefore, the function $v_u(r)$, which equals \,$\mbox{const} \cdot r$\, in the 
thin flow regime \cite{bouch}, now is a constant $v_u$ for $r>0$ and is zero at 
$r=0$.

The physical reasons to introduce equations (\ref{e1}) in such a form are 
described in \cite{bout1,bout2}. One can see that, at a given point, static 
grains converse to rolling ones only if the local slope of the pile is higher 
than the critical angle, and rolling grains converse to static ones, if the 
local slope is lower than the critical angle. 
These linear equations are much simpler for an analytical treatment than the 
previously proposed nonlinear equations for a thin flow regime 
\cite{bouch,bout01,bout02,bout03} in which the characteristic velocity $v_u$ in 
Eq. (\ref{e1}) is replaced by $r(x,t)$ divided by a constant with the 
dimensionality of time. 

In fact, Eqs. (\ref{e1}) represent, maybe, a minimal model for the description 
of granular flows. 
The equations are really very simple: the second equation is independent of the 
first one.
The general solution of Eqs. (\ref{e1}) may be written immediately in the form 
\cite{bout1}:
\begin{eqnarray}
\label{e2}
r(x,t)  & = & u(x+vt) - \frac{v_u}{v + v_u} w(x - v_u t) - v_u \theta_f t,   
\nonumber \\
h(x,t)  & = & w(x-v_u t) + v_u \theta_f t, \;\;\;\;\; r > 0;  \\[7pt]
h(x,t)  & = & \mbox{const}, 
\;\;\;\;\;\;\;\;\;\;\;\;\;\;\;\;\;\;\;\;\;\;\;\;\;\;\;    r = 0, \nonumber
\end{eqnarray}
where $u(x)$ and $w(x)$ are arbitrary functions.
We shall use it to describe the sandpile evolution phenomenologically 
neglecting, as usual \cite{bout1,bout2}, possible near-front deviations from the 
thick flow regime. In fact, in paper \cite{bout2}, moving fronts of granular 
flows were studied in intermediate situations between the thick flow regime and 
the thin flow regime. The results \cite{bout2} show that such a neglect is 
possible.

It is convenient to use also the relation describing conservation of number of 
grains that follows from the continuity equation:

\begin{eqnarray}
\label{e2001}
\int_{-\infty}^0 dx [h(x,t)+r(x,t)] = \nonumber \\
vr_0 t + \int_{-\infty}^0 dx [h(x,0)+r(x,0)] \ .
\end{eqnarray}

It follows directly from Eq. (\ref{e2}) that the solutions consist of the part 
$w(x-v_u t)$ moving with the velocity $v_u$ to the right, the part $u(x+vt)$ 
moving with the velocity $v$ to the left, and the growing (for $h$) or 
decreasing (for $r$) homogeneous background. Specific initial and boundary 
conditions, which we shall use in Secs. II and III, will result in $u(z)$ and 
$w(z)$ consisting only of linear parts. Therefore, the solutions will have 
breaks between linear parts. Some of them will move to the right with the 
velocity $v_u$ and others -- to the left with the velocity $v$.   Coordinates of 
left fronts of the rolling and static grain distributions coincide and, because 
of the decreasing linearly (for $r$) or increasing linearly (for $h$) 
homogeneous background, may move with velocities lower then $v$ if the front of 
$r(x)$ is not of jump-like form.  
One may check, using e.g. Eq. (\ref{e2001}), that a jump-like front of the 
rolling grain distribution $r(x,t)$ may move only with velocity $v$ to the left. 
As it follows from Eq. (\ref{e2}), linear parts of $r(x)$ and $h(x)$ can move 
only parallel to themselves or are motionless. Thus, local slopes of the 
distributions can change only discontinuously if an income flow is constant in 
time.
We shall show the picture of the evolution more clearly for particular examples 
in Secs. II and III.

Thus, to describe the pile evolution we only have to solve linear equations with 
initial and boundary conditions. The boundary conditions are defined from the 
condition on an input flow (we set it constant usually) and from consideration of a moving front of the pile -- a moving boundary. 
In fact, that is the most difficult point in solving problems of this kind. 

There are several possibilities to choose initial conditions. In sections II and 
III we consider two most interesting and natural cases: (i) grains are absent at 
the initial moment, and (ii) there is already a pile with a critical slope 
initially.  We shall show that the pile evolution is a complicated process in 
both cases. 
The slope of a pile approaches  its critical value only at infinite times after 
a long relaxation process during which areas with different slopes are present. At 
long times, the pile evolution looks very similar for both initial conditions, 
although, some distinction can be found even at the infinity. Thus, we observe a 
long memory of initial conditions.   

\begin{figure}
\epsfxsize=75mm
\epsffile{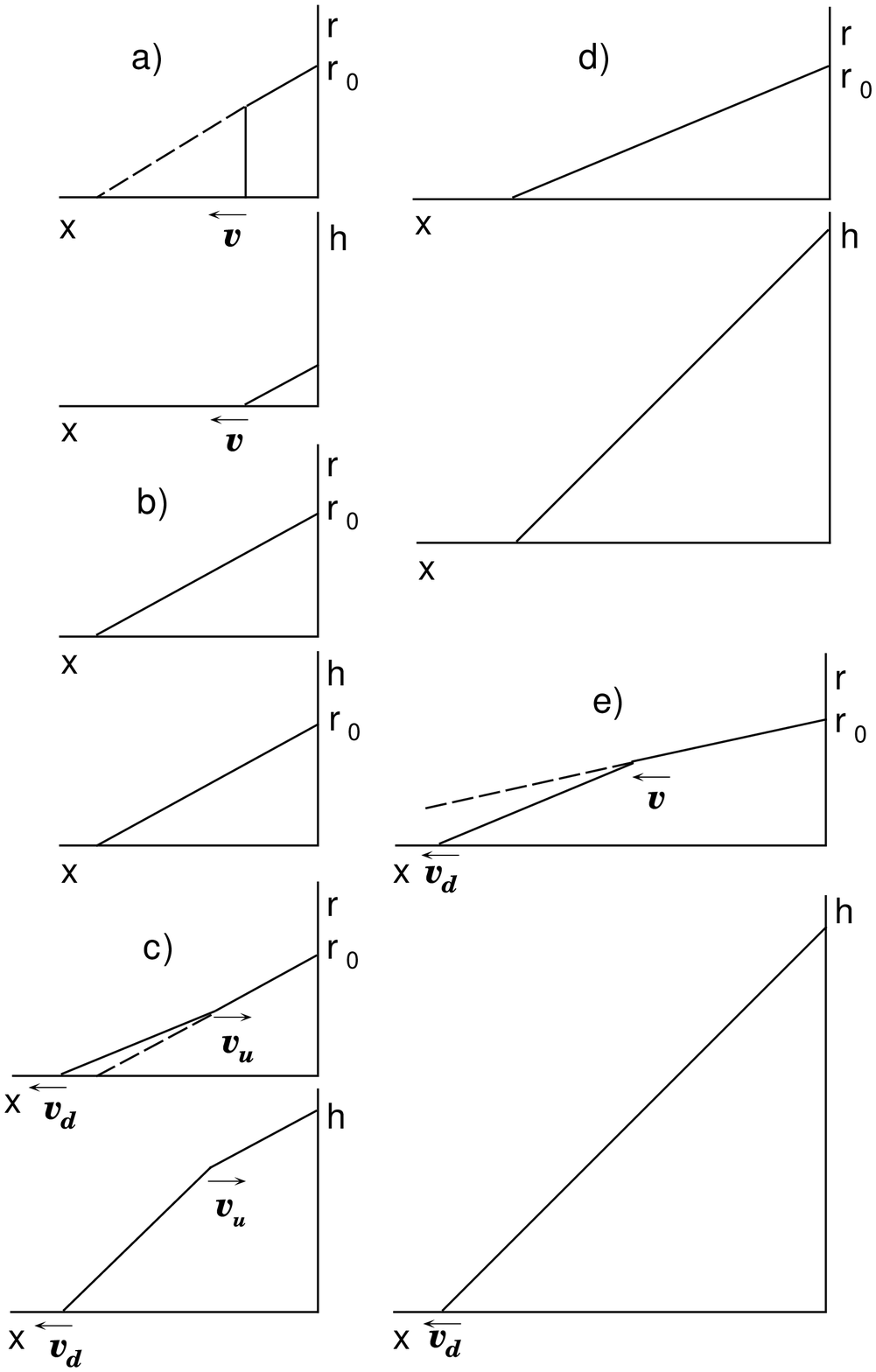}
\caption
{
The evolution of the profiles of rolling grains $r(x)$ and static grains $h(x)$ 
in the region $-\infty<x<0$. In the initial state grains are absent, $r(-
\infty<x<0,t=0)=h(-\infty<x<0,t=0)=0$. 
({\it a})  $0<t<t_1 $, the front moves with the velocity $v$.
({\it b})  $t=t_1 $ [see Fig. 2 and Eq. (4)].
({\it c})  $t_1<t<t_2 $, the front moves with some velocity $v_d<v<v_u$, the 
breaks of the profiles move uphill with the velocity $v_u$.
({\it d})  $t=t_2$, the breaks approach the wall at $x=0$.
({\it e})  $t_2<t<t_3$, the front proceeds to move with the velocity $v_d$, the 
break of $r(x)$ moves downhill with the velocity $v$. Note, that the right 
linear part of $r(x)$ is always motionless. After the break of $r(x)$ overtakes 
the front at $t=t_3$, the general configurations ({\it b}) -- ({\it e})  
is repeated with a lower front velocity.
}
\label{fig1}
\end{figure}

The results obtained for initial conditions (i) and (ii) are presented in Figs. 
\ref{fig1} and \ref{fig5} -- movies of $r$ and $h$ distribution profiles, in 
Figs. \ref{fig2} and \ref{fig6} -- diagrams of trajectories of fronts and breaks 
of the grain distributions on the $t,x$-plane, and in Figs. \ref{fig3} and 
\ref{fig7} -- the dependencies of the pile slopes on time.

In fact, all these figures depict a long relaxation process to the critical 
state (i.e. to the pile with the critical angle) by different means. For 
example, from Fig. 3 one sees that the relaxation is discontinuous -- local 
derivatives of the slope over time are zero or infinity for all times.  

We shall demonstrate the space periodicity appearing while a pile evolves: it 
turns out that grains are piling layer by layer (see Fig. \ref{fig2} and \ref{fig6}) 
and the thickness of the layers just coincides with the thickness of the 
stratified layers observed in two-component sand-piles  \cite{makse97prl,ciz}! 
Thus, we give positive answer to the question stated above and find a precursor 
of an intriguing stratification phenomena already in an one-component sandpile 
problem.

One may wonder now if there is any possibility for a sandpile to evolve without 
a space periodicity at all. We demonstrate that, of course, such evolution is 
possible in a thick flow regime, for example, if an input flow is a linear 
function of time. We show that the front moves with a constant velocity all the 
time, and linear profiles of moving and static grains expand without any 
surprises in that situation (see Fig. \ref{fig4}). The relation between amounts of 
static and rolling grains in a pile depends on the rate of the input flow 
increase.

\section{Sandpile evolution starting from empty state}

\subsection{Constant input flow}

The case we will consider in this section corresponds to the situation when 
there are no particles in the initial state.
 
Let there be a wall at $x=0$, and grains be poured permanently beginning from 
the moment $t=0$ at this point, so $r(x=0, t)=r_0$ is a boundary condition 
($r_0$ is the thickness of the input flow). 
Let, first, $r_0$ be constant in time. The sandpile is supposed to expand to the 
left, i.e. to $x = -\infty$. There are no particles  at the initial stage so $r(x\le 
0, t=0) =0$, $h(x\le 0, t=0) =0$ are the initial conditions. Inserting general 
solutions from Eqs. (\ref{e2}) and the above initial and boundary conditions one 
may obtain the functions $u(z)$ and $w(z)$. 

To clear up the picture and to simplify the calculations
we start from the following consideration of the begininning stage. 
(Below, we shall demonstrate consistent complete calculations in 
the more frequent case, which is 
realized at more long times, of a $r(x)$ front linearly decreasing to zero.) 

One may see that because of the initial and boundary conditions for $r(x,t)$ and 
of the form of general solution for it Eq. (\ref{e2}), $r(x,t)$ should first has a 
jump front; the part of $r(x,t)$ adjacent to $x=0$ has to be independent of $t$. 
The last statement follows immediately from the form of the general solutions 
Eq. (\ref{e2}) and from the fixed right boundary condition  $r(x=0, t)=r_0$.  
The first statement follows from the last one -- the jump front of $r(x,t)$ is 
the only possibility for the first stage of the process. But then, as we have 
seen in Sec. I, it has to move with velocity $v$. 

The distribution of static grains $h(x,t)$ which appears from  the precipitation 
of rolling ones has a front that goes to zero without an abrupt leap but 
linearly in the case under consideration, since there are no fixed boundary 
conditions for $h$ in the right. Then one can easily imagine that at first the 
solutions have to look as shown in Fig. 1(a) (as we have noticed, nothing but 
linear functions may appear from the considered initial and boundary 
conditions). In principle, one may immediately insert $r$ and $h$ in this form 
(\,$r(x,t)=r_0-a x, \ h(x,t)=b(x+vt)\,$, where $a$ and $b$ are constant 
coefficients) into Eq. (\ref{e1}) and obtain the following answer in the time 
interval $0<t<t_{1} = (v+v_u)r_0/(v v_u \theta_f)$:

\begin{eqnarray}
\label{e3}
r(x,t) & = & \left(r_0 + \frac{v_u}{v+v_u} \; \theta_{f} x \right) \Theta(x+vt)  
\nonumber \\[3pt]
h(x,t) & = & \frac{v_u}{v+v_u} \; \theta_{f}(x+vt)  
\end{eqnarray}
$\Theta(x)$ is the Heaviside function (we do not write the multiplier 
$\Theta(x+vt)$ in the right hand side part of the second equation of Eqs. 
(\ref{e3}), since, of course, $r(x)$ and $h(x)$ can not be negative). The static 
grains are precipitated from the rolling ones, so the front coordinates of 
$r(x)$ and $h(x)$ should coincide. The meaning of the time $t_1$ is clear from 
Fig. 1(a) -- it is the very first instant at which the jump at the front appears 
to be zero. Note, that the slope of the static part $\partial h/\partial x = v_u 
\theta_f/(v+v_u)$ is less than the critical slope, so the relaxation to the 
critical slope is non trivial.

Nevertheless, to be sure, we prefer to substitute the general solutions Eq. 
(\ref{e2}) into the following set of initial and boundary conditions

\begin{eqnarray}
\label{e3000}
r(x,t=0) & = & 0 \ ,   \nonumber \\
h(x,t=0) & = & 0 \ ,   \nonumber \\
r(x=0,0<t<t_1) & = & r_0 \ ,   \nonumber \\
h(x=-vt,0<t<t_1) & = & 0 \ .   
\end{eqnarray}

Using previous considerations to simplify the calculations, we assume in Eq. 
(\ref{e3000}) the front velocity to be equal to $v$ from the very beginning, so a 
fifth condition for $r(x,t)$ at the front point is not necessary. In principle, 
the time $t_1$ may be obtained if one demands that all parts of functions 
$u(z)$ and $w(z)$ have to be connected together continuously, but we prefer to use the 
already known expression for $t_1$ and be sure that calculations are 
self-consistent (i.e. all parts of functions $u(z)$ and $w(z)$ are connected together 
continuously) only afterwards. 

Substituting the general solutions Eq. (\ref{e2}) into Eq. (\ref{e3000}) we get

\begin{eqnarray}
\label{e2002}
w\left(-\frac{(v+v_u)^2}{v v_u}\frac{r_0}{\theta_f}<z<0\right) & = &  
\frac{v_u}{v+v_u}\theta_f z \ , \nonumber \\[3pt]
u\left(0<z<\frac{v+v_u}{v_u}\frac{r_0}{\theta_f}\right) & = & 
r_0+\frac{v_u(v+2v_u)}{(v+v_u)^2}\theta_{f} z .
\end{eqnarray}

After substitution Eq. (\ref{e2002}) into Eq. (\ref{e2}) we again obtain our 
solution Eq. (\ref{e3}) 
of Eq. (\ref{e1}) with a front moving to the left with the velocity $v$ in the 
time interval $0<t<t_1$ [see Fig.1(a)]. 

As a result, at the time $t_1$ one gets

\begin{mathletters}
\label{e4}
\begin{equation}
r(x,t_1) = r_0 + \frac{v_u}{v+v_u}\theta_f x , \label{e4a}
\end{equation}
\begin{equation}
h(x,t_1) = r_0 + \frac{v_u}{v+v_u}\theta_f x \label{e4b}
\end{equation}
\end{mathletters}
for $-[(v+v_u )/(vv_u )]r_0/\theta_f<x<0$ [Fig. 1(b)]. These equations are used 
as initial conditions to find the solutions in the next time interval $t_1 < 
t < t_3$, at which the front of $r(x,t)$ will be jumpless. 

Times $t_2$ and $t_3$ appear naturally from the solution (see Fig. 1)
but we prefer to write out their expressions immediately: 
$t_{2}=[(v+v_u)^2/v v_u^2]r_0/\theta_f$ and $t_3=[(v+v_u)(v+3v_u)/v 
v_u^2]r_0/\theta_f$. 
As one may understand from the figure, the meaning of the times is the 
following: $t_2$ is the time at which breaks of $r(x,t)$ and $h(x,t)$ moving 
from the front to the right with the velocity $v_u$ will approach the point 
$x=0$; and $t_3$ is the time at which the break of $r(x,t)$ moving from $x=0$ 
will overtake the front. 

To simplify our calculations, we again use the expression for $t_3$ and  
check the correctness of the choice of $t_3$ at the very end of the 
calculations.

Now we show how to treat moving boundary conditions. 
We have to add to the initial conditions Eq. (\ref{e4}) the following boundary 
conditions:

\begin{mathletters}
\label{e5}
\begin{equation}
r(x=0,t>0) =  r_0 , \label{e5a}
\end{equation}
\begin{equation}
r\left(x=-\frac{v+v_u }{vv_u }\frac{r_0}{\theta_f}-v_d(t-t_1),t_1<t<t_3\right)   
=  0, \label{e5b}
\end{equation}
\begin{equation}
h\left( x=-\frac{v+v_u }{vv_u }\frac{r_0}{\theta_f}-v_d(t-t_1),t_1<t<t_3 \right)   
=  0. \label{e5c}
\end{equation}
\end{mathletters}

Eqs. (\ref{e5b}) and (\ref{e5c}) are the conditions for the left front of the 
pile that is supposed to move with a yet unknown velocity $v_d$. All we have to 
do is (i) to insert the general solutions Eq. (\ref{e2}) into Eqs. (\ref{e4}) 
and (\ref{e5}); (ii) to find  $v_d$, $u(z)$, and $w(z)$; and (iii) to check the 
self-consistency of our choice of $t_3$ which was made beforehand, in fact, from 
physical reasons. Note that the beforehand choice of $t_3$ was used only to 
make the following intermediate formulas more compact. 

Substitution of Eq. (\ref{e2}) into Eq. (\ref{e5c}), (\ref{e4b}), (\ref{e4a}), 
and (\ref{e5b}) gives

\begin{eqnarray}
\label{e5001}
w( -\frac{v+v_u}{v v_u^2}(v_d v + 2v_u v + 2v_d v_u + 
3v_u^2)\frac{r_0}{\theta_f}<z<
\nonumber \\ 
-\frac{(v+v_u)^2}{v v_u}\frac{r_0}{\theta_f}) = \frac{(v+v_u)(v-
v_d)}{v(v_u+v_d)}r_0+\frac{v_u}{v_u+v_d}\theta_f z ,
\end{eqnarray}

\begin{equation}
\label{e5002}
w\left(-\frac{(v+v_u)^2}{v v_u}\frac{r_0}{\theta_f}<z<-
\frac{v+v_u}{v}\frac{r_0}{\theta_f}\right) = \frac{v_u}{v+v_u}\theta_f z ,
\end{equation}

\begin{equation}
\label{e5003}
u\left(0<z<\frac{v+v_u}{v_u}\frac{r_0}{\theta_f}\right) = r_0 + 
\frac{v_u(v+2v_u)}{(v+v_u)^2}\theta_f z ,
\end{equation}
and

\begin{eqnarray}
\label{e5004}
u\left(0<z<\frac{(v-v_d)(v+v_u)(v+2v_u)}{v v_u^2}\frac{r_0}{\theta_f}\right) =  
\nonumber \\[2pt]
r_0 + \frac{v v_u}{(v-v_d)(v+v_u)}\theta_f z 
\end{eqnarray}
correspondingly.
The functions $u(z)$ from Eqs. (\ref{e5003}) and (\ref{e5004}) have to coincide 
since they are defined at (at least) overlapping intervals of $z$ (we shall see 
below that, in fact, these intervals coincide). Thus, equating Eqs. 
(\ref{e5003}) and (\ref{e5004}) we obtain the answer for $v_d$ in the case under 
consideration:

\begin{equation}
\label{e5005}
v_d = \frac{v v_u}{(v+2v_u)} .
\end{equation}
One sees immediately that $v_d<v,v_u$. Inserting this expression into Eq. 
(\ref{e5004}) we find that the intervals of the variable $z$ for $u(z)$ in Eqs. (\ref{e5003}) 
and (\ref{e5004}) are the same.

Inserting Eq. (\ref{e5005}) into Eq. (\ref{e5001}) we obtain

\begin{eqnarray}
\label{e5006}
w\left(-3\frac{(v+v_u)^2}{v v_u}\frac{r_0}{\theta_f}<z<-\frac{(v+v_u)^2}{v 
v_u}\frac{r_0}{\theta_f}\right) =  \nonumber \\[2pt]
\frac{v+v_u}{2v_u}r_0 + \frac{(v+2v_u)}{2(v+v_u)}\theta_f z .
\end{eqnarray}
At last, substitution of Eq. (\ref{e2}) into Eq. (\ref{e5a}) gives, accounting 
for already known answers for $w(z)$ Eqs. (\ref{e5002}) and (\ref{e5006}) in 
different intervals of $z$, the following expressions for $u(z)$:

\begin{eqnarray}
\label{e5007}
u\left(\frac{(v+v_u)}{v_u}\frac{r_0}{\theta_f}<z<\frac{(v+v_u)^2}{v_u^2}\frac{r_
0}{\theta_f}\right) = \nonumber \\[2pt]
r_0 + \frac{v_u(v+2v_u)}{(v+v_u)^2}\theta_f z
\end{eqnarray}
and
\begin{eqnarray}
\label{e5008}
u\left(\frac{(v+v_u)^2}{v_u^2}\frac{r_0}{\theta_f}<z<3\frac{(v+v_u)^2}{v_u^2}
\frac{r_0}{\theta_f} \right) = 
\nonumber \\[2pt]
\frac{3}{2}r_0 + \frac{v_u(2v+3v_u)}{2(v+v_u)^2}\theta_f z .
\end{eqnarray}

\begin{figure}
\epsfxsize=80mm
\epsffile{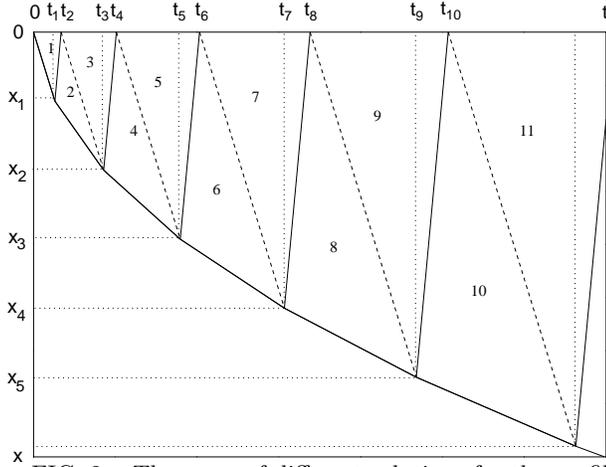}
\caption{
The areas of different solutions for the profiles of rolling and static grains 
for the pile evolution starting from an empty state [see Eqs. (\protect\ref{e9}) 
and (\protect\ref{e10})]. $v/v_u=0.3$\,. The lower segmented line shows the 
dependence of the front coordinate on time. The solid lines $x=v_u(t-t_{2m}), 
m=1,2,\ldots$ depict the uphill movement of the $r(x)$ and $h(x)$ breaks. The 
dashed lines $x=-v(t-t_{2m})$ show the downhill movement of the break of the 
$r(x)$ profile. The points $x_m$ are arranged periodically.}
\label{fig2}
\end{figure}

\begin{figure}
\epsfxsize=70mm
\epsffile{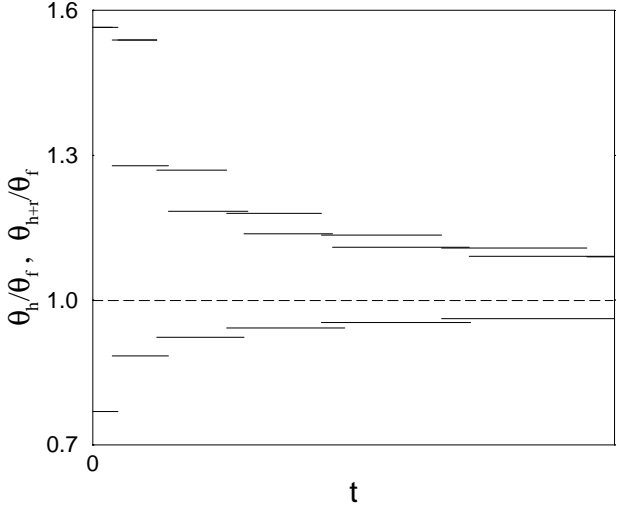}
\caption{
The dependence of the relative pile slope on time when the pile evolution starts 
from an empty state. $v/v_u=0.3$\,. $\theta_h$ is the slope of a static part of 
the pile. $\theta_{h+r}$ is the slope of the whole pile consisting of static and 
rolling parts. The upper set of lines shows $\theta_{h+r}$. At infinity 
$\theta_h, \theta_{h+r} \to \theta_f$. The separated lines for $\theta_{h+r}$ are 
defined for $0<t<t_2$, $t_1<t<t_3$, $t_2<t<t_4$, etc. (see Fig. 2).  The lines 
for $\theta_h$ are defined for $0<t<t_2$, $t_1<t<t_4$, $t_3<t<t_6$, etc. 
Sometimes (e.g., at $t_1<t<t_2$, $t_3<t<t_4$, etc. for $\theta_h$) the profiles 
have two parts with different slopes.}
\label{fig3}
\end{figure}

\begin{figure}
\epsfxsize=85mm
\epsffile{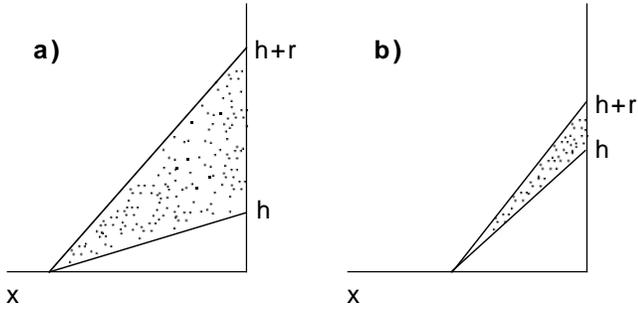}
\caption{
The evolution of a sandpile with a linearly growing input flow: $r_0 = v_0 t$, 
$v_0=\mbox{const}$. An initial state is empty. We show here both static and 
rolling grains in the same plots, so profiles for $h(x)$ and $h(x)+r(x)$ are 
presented. The fronts moves with a time-independent velocity $v_d$ which is a 
function of $v_0$.
({\it a}) The rate $v_0$ is much higher than 
$v_0^\ast$ [see Eq. (\protect\ref{i5}) and the text]. The front velocity is 
close to its maximal possible value $v$, rolling grains have no time to converse 
into static ones, so the amount of rolling grains is much greater than the 
amount of static grains.  
({\it b}) The rate $v_0$ is much lower than $v_0^\ast$. A relative amount of 
rolling grains is small, the front velocity tends to zero, and the slope of the 
distribution of static grains is nearly critical. 
\label{fig4}
}
\end{figure}

\begin{figure}
\epsfxsize=160mm
\epsffile{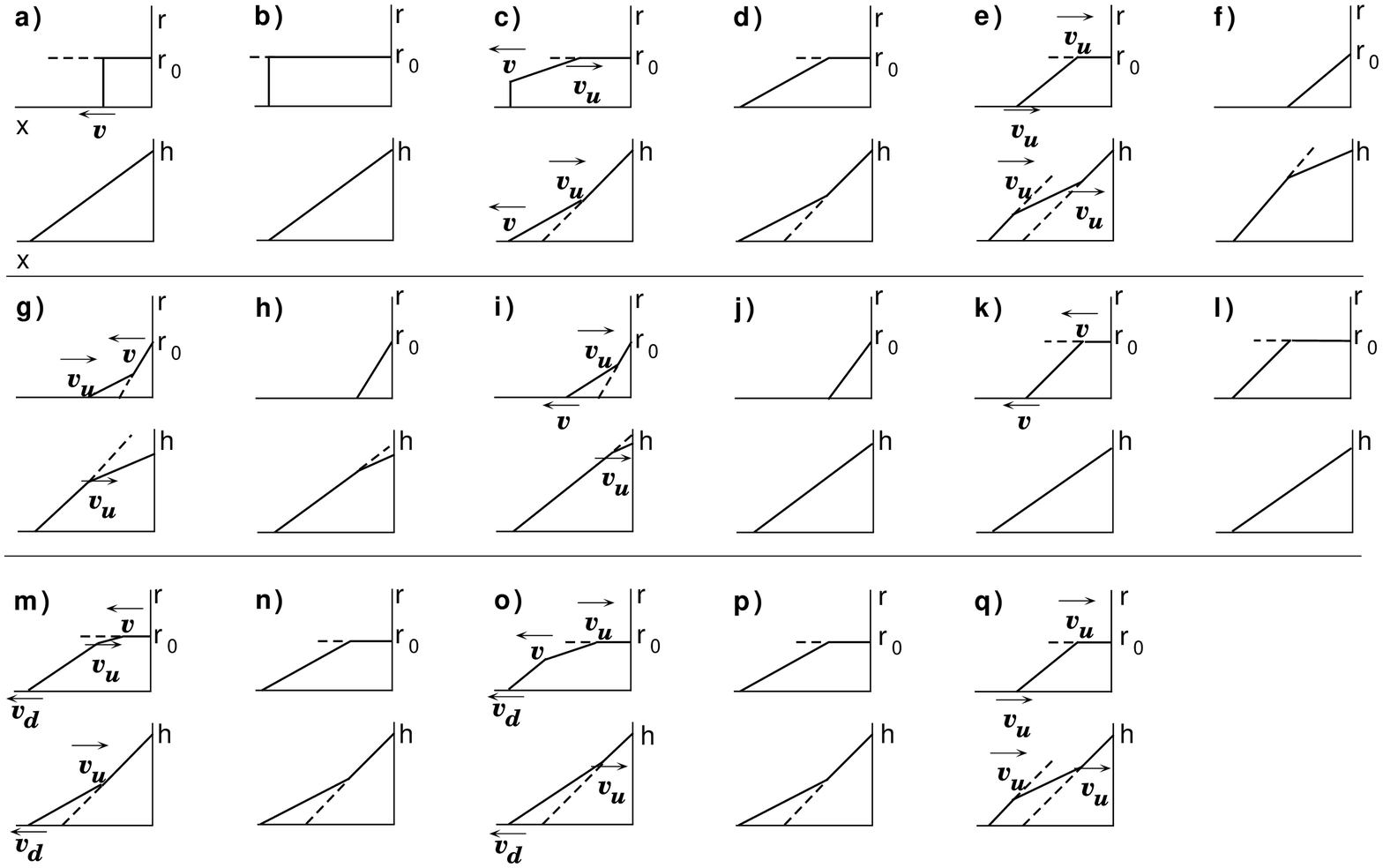}
\caption{
The evolution of the profiles of rolling grains $r(x)$ and static grains $h(x)$ 
starting from the critical state (see the text for details). At the initial moment 
$r(-\infty<x<0,t=0)=0$ and $h(-\infty<x<0,t=0)=\theta_f(x+d)$. Plots 
({\it a})--({\it q}) show the profiles at some successive instants. 
Afterwards the evolution 
proceeds in a similar way. The scale is changed from one plot to another.
\label{fig5}
}
\end{figure}

\begin{figure}
\epsfxsize=80mm
\epsffile{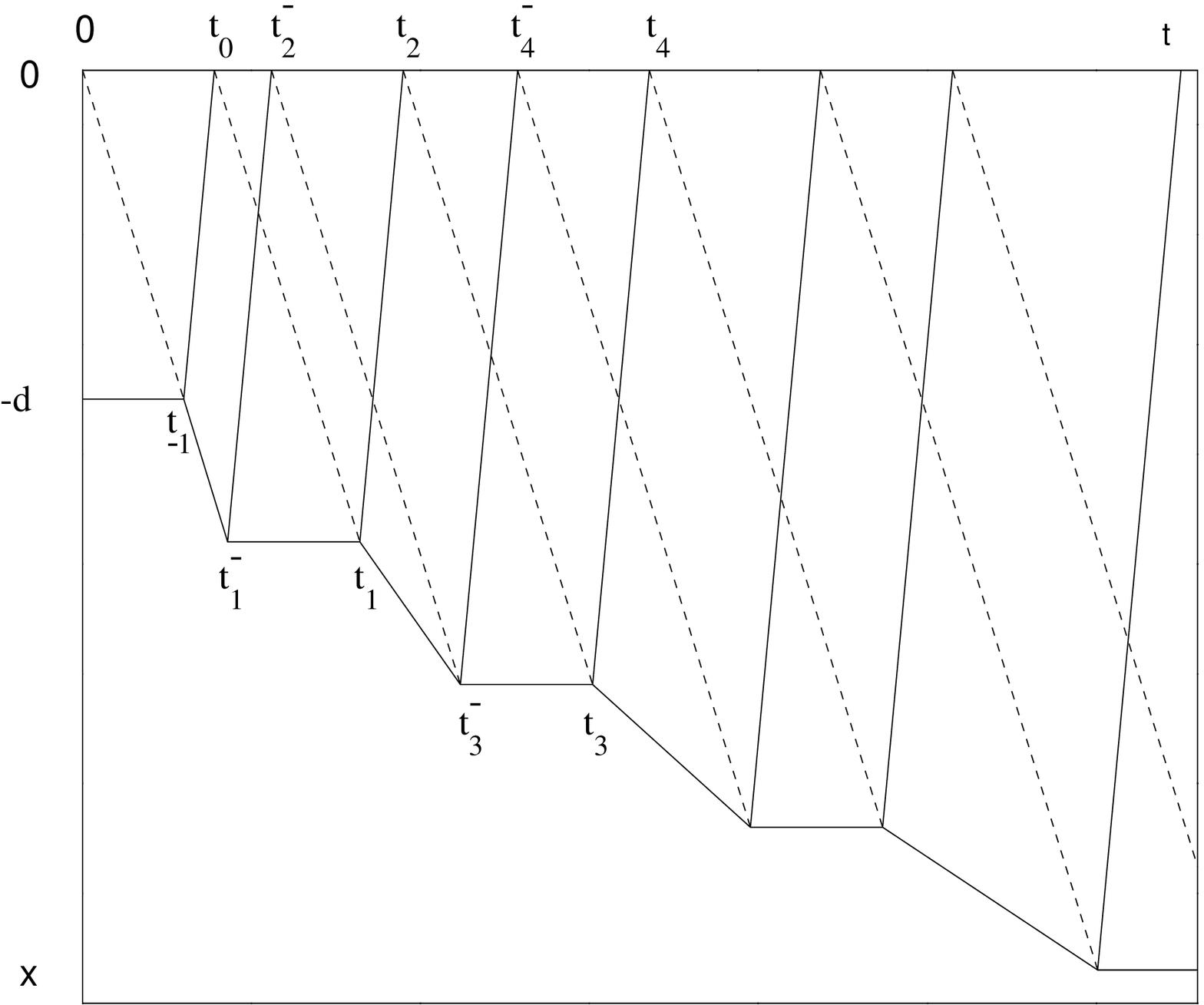}
\caption{
Trajectories of the front and the breaks of the $r(x)$ and $h(x)$ profiles on 
$t,x$-plane in the case of the pile evolution starting from a critical state 
(compare with Fig. 2). $v/v_u=0.3$\,. 
$h(x,t=0)=\theta_f(x+d)$. The lower segmented line shows the dependence of the 
front coordinate on time. The solid lines $x=v_u(t-t_{2m}), m=1,2,\ldots$ depict 
the uphill movement of the $r(x)$ and $h(x)$ breaks. The dashed lines $x=-v(t-
t_{2m})$ show the downhill movement of the break of the $r(x)$ profile. The 
coordinates of plateaus are arranged periodically. 
\label{fig6}
}
\end{figure}

\begin{figure}
\epsfxsize=70mm
\epsffile{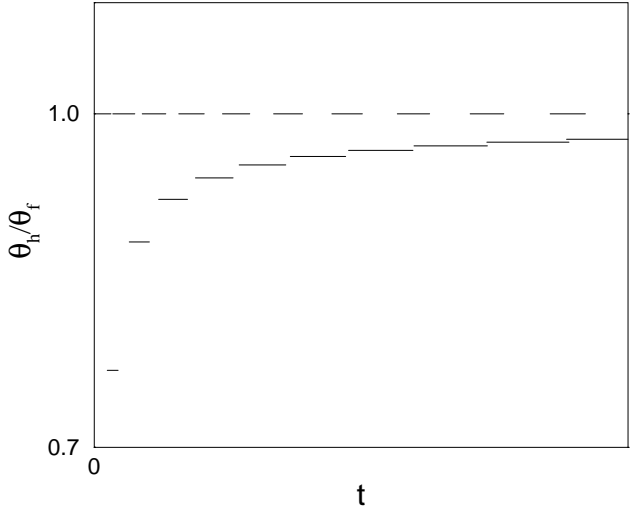}
\caption{
The dependence of the relative pile slope $\theta_h/\theta_f$ on time when the 
pile evolution starts from the critical state (compare with Fig. 3). 
$v/v_u=0.3$\,. In some time intervals the profile of static grains has two or even three 
parts with different slopes. Segments with the slope equal exactly to the 
critical one are present even at long times.
\label{fig7}
}
\end{figure}

Combining Eqs. (\ref{e5003}) and (\ref{e5007}) we get 
\begin{equation}
\label{e5009}
u\left(0<z<\frac{(v+v_u)^2}{v_u^2}\frac{r_0}{\theta_f}\right) = 
r_0 + \frac{v_u(v+2v_u)}{(v+v_u)^2}\theta_f z .
\end{equation}
Eqs. (\ref{e5008}), (\ref{e5009}), (\ref{e5002}), and (\ref{e5006}) present a 
full answer for $u(z)$ and $w(z)$. One may see easily that they are bounded 
continuously, so our choice of $t_3$ was correct. Inserting Eqs. (\ref{e5008}), 
(\ref{e5009}), (\ref{e5002}), and (\ref{e5006}) into Eq. (\ref{e2}) we obtain a 
full solution of the problem for $t_1<t<t_3$: 
\end{multicols}
\widetext
\noindent\rule{20.5pc}{0.1mm}\rule{0.1mm}{1.5mm}\hfill
\begin{mathletters}
\label{e5010}
\begin{equation}
r(v_u(t-t_{2})<x<0,t>t_{1})  =  r_0 + \frac{v_u}{v+v_u}\,\theta_f x , 
\label{e5010a}
\end{equation}
\begin{equation}
r\left(-\frac{(v+v_u)^2}{v_u(v+2v_u)}\frac{r_0}{\theta_f}-\frac{v 
v_u}{v+2v_u}t<x<
-v(t-t_{2}), v_u(t-t_2)\right)  =  
\frac{r_0}{2} + \frac{v_u(v+2v_u)}{2(v+v_u)^2}\,\theta_f \left(x+\frac{v 
v_u}{v+2v_u}t\right) , 
\label{e5010b}
\end{equation}
\begin{equation}
r(-v(t-t_{2})<x<0,t<t_{3})  =  r_0 + \frac{v_u}{2(v+v_u)}\,\theta_f x ,
\label{e5010c}
\end{equation}
\begin{equation}
h(v_u(t-t_{2}),-vt<x<0)  =  \frac{v_u}{(v+v_u)}\,\theta_f (x+vt) , 
\label{e5010d}
\end{equation}
\begin{equation}
h\left(-\frac{(v+v_u)^2}{v_u(v+2v_u)}\frac{r_0}{\theta_f}-\frac{v 
v_u}{v+2v_u}t<x<
-v_u(t-t_{2}),0,t<t_{3}\right)  =   
\frac{v+v_u}{2v_u}r_0 + \frac{(v+2v_u)}{2(v+v_u)}\,\theta_f \left(x+\frac{v 
v_u}{v+2v_u}t \right) . 
\label{e5010e}
\end{equation}
\end{mathletters}
\hfill\rule[-1.5mm]{0.1mm}{1.5mm}\rule{20.5pc}{0.1mm}
\begin{multicols}{2}
\narrowtext
These solutions are shown in Fig. 1(b--e) [see also Fig. \ref{fig2}, the region 
$t_1<t<t_3$ in which trajectories of the front -- a lower segmented line -- and 
the breaks of $r(x,t)$ and $h(x,t)$ are depicted]. The solutions (\ref{e5010a}) 
and (\ref{e5010d}) are defined in region $1$ of Fig. \ref{fig2}, the solution 
(\ref{e5010b}) is defined in region $2$ of the figure, the solution 
(\ref{e5010c}) is defined in region $3$, and the solution (\ref{e5010e}) for 
$h(x,t)$ is defined in both regions $2$ and $3$. 

The evolution looks as the following. At the instant $t_1$ new linear parts of 
$r$ and $h$ appear at the front point $x=-[(v+v_u )/v_u ]r_0/\theta_f$. The new 
front moves to the left with the velocity $v_d$, breaks of $r(x,t)$ and $h(x,t)$ 
move to the right with the velocity $v_u$ -- a solid line $x=v_u(t-t_2)$ in Fig. 
2. The slope of this part of the static grain distribution $h(x,t)$ is higher 
than it was at the first stage but lower than the critical slope. The breaks 
approach the $x=0$ point at $t=t_2$. Then a new time-independent 
linear part of $r(x)$ appears close to the wall. The left part 
proceeds to move to the left, so we see a break moving downhill -- a dashed line 
$x=-v(t-t_2)$ in Fig. 2. Its velocity equals $v$. Note that, unlike $r(x,t)$, 
$h(x,t)$ has no breaks for $t_2<t<t_3$. At the instant $t_3$ the break of $r(x,t)$ 
overtakes the front. That is possible, since the velocity of the front is lower 
than the downhill velocity of the break, $v_d<v$.

Therefore we confirm the general picture of the evolution predicted in Sec. I: 
there exist breaks of both $r(x,t)$ and $h(x,t)$ moving uphill with the velocity 
$v_u$, and only the break of $h(x,t)$ is moving downhill with the velocity $v$. 
When the breaks approach a wall or the break of $h(x,t)$ overtakes the front, 
new linear parts of the solutions appear. 

From our solutions Eq. (\ref{e5010}), we obtain new initial conditions for 
$r(x,t)$ and $h(x,t)$ at the moment $t=t_3$. Then we may repeat the described 
procedure for the next time interval $t_3 < t < t_5 $, etc. However, now, when 
we understand the structure of the solution, one may again simplify the 
calculations. Instead of calculating the functions $u(z)$ and $w(z)$ one may 
proceed directly with $r(x,t)$ and $h(x,t)$. Areas of different linear parts of 
solutions are triangular regions in Fig. \ref{fig2} separated one from each 
other by solid or dashed lines -- trajectories of the breaks. Let us suppose that 
the solution is known in the region $(0,0)-(x_1,t_1)-(0,t_2)$, i.e. in region $1$ in 
Fig. \ref{fig2}. One may sew easily an unknown linear solution in the adjusted 
triangular region $2$ together with the known one along the line $x=-v(t-t_2)$. 
It is easy to find coefficients of linear terms and, therefore, the front 
velocity. 
From this values one finds the shape $(x_1,t_1)-(0,t_2)-(x_2,t_3)$ of region $2$ 
in which the new part of the solution is defined (see Fig. \ref{fig2}). Then we 
repeat the described procedure for the next triangle, etc.

After these simple but rather tedious calculations we obtain the total solution 
consisting of linear parts, a structure of which one can see from  Figs. 1 and 
\ref{fig2} and following equations. The lowest segmented line in Fig. \ref{fig2} 
shows the dependence on time of the front position. Coordinates of the segments 
are:
\begin{eqnarray}
\label{e6}
x(t_{2m-1}<t<t_{2m+1}) = -\frac{m(m+1)}{2} \frac{(v+v_u)^{2}}{v_{u} {\cal 
V}_{m}} \frac{r_0}{\theta_f} \nonumber \\[3pt]
- \frac{v v_u}{{\cal V}_{m}} t, \;\;\; m=0,1,2,\dots
\end{eqnarray}
(for $m=0$ the time interval is $0<t<t_{1}$) with ${\cal V}_{m} \equiv mv + 
(m+1)v_u $. The particular times shown in Fig. 2 are

\begin{eqnarray}
\label{e7}
t_{2m-1} & = & \frac{m}{2} \frac{(v+v_u)[(m-1)v+(m+1)v_{u}]}{v v_{u}^2} 
\frac{r_0}{\theta_f} , \nonumber \\[3pt]
t_{2m}   & = & \frac{m(m+1)}{2} \frac{(v+v_u)^2}{v v_{u}^2} \frac{r_0}{\theta_f} 
,  \\[5pt]
m        & = &1,2,\dots   \nonumber
\end{eqnarray}
The front coordinates corresponding to times $t_{2m-1}$ at which the front 
velocity changes its value are

\begin{equation}
\label{e8}
x(t_{2m-1}) = -m \frac{(v+v_u)}{v_{u}} \frac{r_0}{\theta_f} \ , \ m=1,2\ldots
\end{equation}
Hence, these points are arranged periodically. Two other types of lines are 
shown in Fig. 2: solid lines $x=v_u (t-t_{2m})$ and dashed lines $x=-v(t-
t_{2m}), m=1,2,\ldots$ The lines of the first type depict the uphill movement of 
the breaks of both profiles $r(x)$ and $h(x)$ 
[Fig.1(c)]. The lines of the second type show the downhill movement of 
the break of the profile $r(x)$ with the velocity $v$ [see Fig.1(e)]. 

Solutions for all regions of Fig. 2, connected at these lines, look as:
\end{multicols}
\widetext
\noindent\rule{20.5pc}{0.1mm}\rule{0.1mm}{1.5mm}\hfill
\begin{eqnarray}
\label{e9}
r\left( -v(t-t_{2m-2}),v_u (t-t_{2m})<x<0 \right)  =  r_0 + \frac{v_u}{m(v+v_u)} 
\; \theta_f x, \nonumber \\
r\left(-\frac{m(m+1)}{2}\frac{(v+v_u)^2}{v_u {\cal V}_{m}} \frac{r_0}{\theta_f}-
\frac{v v_u} {{\cal V}_{m}} t<x< -v(t-t_{2m}),v_u (t-t_{2m}) \right)  =  
\frac{r_0}{2}+ \frac{1}{m(m+1)}\frac{v_u {\cal V}_{m}}{(v+v_u)^2} \; \theta_f 
\left(x+\frac{vv_u} {{\cal V}_{m}} t \right), \\[3pt]
m         = 1,2,\dots   \nonumber 
\end{eqnarray}
\begin{equation}
\label{e10}
h(v_u(t-t_{2m+2})<x<v_u (t-t_{2m}),0)  = \frac{m}{2}\frac{v+v_u}{v_u}r_0 + 
\frac{{\cal V}_{m}} {(m+1)(v+v_u)} \; \theta_f \left(x+\frac{vv_u}{{\cal 
V}_{m}}t \right), \;\;\; m = 0,1,2,\dots  
\end{equation}
\hfill\rule[-1.5mm]{0.1mm}{1.5mm}\rule{20.5pc}{0.1mm}
\begin{multicols}{2}
\narrowtext
\noindent
Here, we set $t_0=0$. Inequalities in the right hand parts of Eqs. (\ref{e9}) 
and (\ref{e10}) define the areas of validity of the solutions (see Fig. 2).
Eqs. (\ref{e9}) and (\ref{e10}) describe totally the sandpile evolution, see 
Fig. 1(b-e).

From Eq. (\ref{e10}) and Fig. 2, a space periodicity of the process is evident: 
general shapes of the profiles $r(x)$ and $h(x)$ are repeated each time the 
front moves by $(v+v_u)r_0/(v_u \theta_f)$ to the left. In fact, 
the pile is increased layer by layer, and  the expression for the width of these 
layers is the same as the one for the width of stratified layers of different 
fractions in a two-component sandpile (that was obtained in papers 
\cite{makse97prl,ciz}). One sees from Eq. (\ref{e10}) that heights of the pile at 
times $t_{2m}$ are also periodic in $m$:

\begin{equation}
\label{e11}
h(x=0,t_{2m}) = m \; \frac{v+v_u}{v_u} \; r_0 \ , m=0,1,2\ldots
\end{equation}

It follows from Eq. (\ref{e10}) that the slope, $\theta_h \equiv \partial h / 
\partial x$, of the static part of the pile will approach its critical value 
only at infinite time:

\begin{equation}
\label{e12}
\theta_h(t_{2m-3}<t<t_{2m}) = \left(1- \frac{v}{m\,(v+v_u)}\right)\theta_f \ , 
m=1,2\ldots
\end{equation}
(for $m=1, \ 0<t<t_2$) (see Fig. 3). At $t_{2m-1}<t<t_{2m}, m=1,2,\ldots$ there 
are two different slopes for two parts of the profile $h(x)$. For 
$t_{2m}<t<t_{2m+1},m=0,1,2,\ldots$ all the profile has the same slope. Thus, for 
long times $t\gg[(v+v_u)^2/(vv_u^2)]r_0/\theta_f$ the slope behaves as

\begin{equation}
\label{e13}
\theta_h \cong \left(1-\frac{v}{v_u} \sqrt{\frac{r_0}{2v\,\theta_f 
t}\,}\right)\theta_f 
\ , 
\end{equation}
and $\theta_h$ relaxes slowly to its final value $\theta_f$ by a power law.

A slope of the whole pile -- including both static and rolling parts, 
$\theta_{h+r} \equiv \partial (h+r) / \partial x$, depends on time in the 
following way:
 
\begin{eqnarray}
\label{e131}
\theta_{h+r}(t_{2m-2}<t<t_{2m}) & = & \left(1 + \frac{v_u-
v}{m\,(v+v_u)}\right)\theta_f \ , 
\nonumber \\[3pt]
\theta_{h+r}(t_{2m-1}<t<t_{2m+1}) & = &
\nonumber \\[3pt]
 &  &  
\!\!\!\!\!\!\!\!\!\!\!\!\!\!\!\!\!\!\!\!\!\!\!\!\!\!\!\!\!\!\!\!\!\!\!\!\!\!\!\!
\!\!\!\!\!\!\!\!\!\!\!\!\!\!\!\!\!\! \left(1 + \frac{v_u-v}{m\,(v+v_u)} + 
\frac{v^2}{m(m+1)(v+v_u)^2} \right)\theta_f \ , 
\nonumber \\[3pt]
 m=1,2\ldots
\end{eqnarray}
(see Fig. \ref{fig3}) and approaches the final value $\theta_f$ from above at 
long times.
It behaves asymptotically as

\begin{equation}
\label{e132}
\theta_{h+r} \cong \left(1+\frac{v_u-v}{v_u} \sqrt{\frac{r_0}{2v\,\theta_f 
t}\,}\right)\theta_f  
\end{equation}
[compare with Eq. (\ref{e13})].

 At long times, the coordinate of the front is $x\cong\sqrt{2 v r_0 t/\theta_f}$, 
and its velocity tends to $-\sqrt{v r_0/(2\theta_f t)}$. Two last relations are 
obvious. Indeed, at long times the slope of a pile is close to its critical 
value. The relative part of rolling grains is small.
After substitution of a linear function with the critical slope into the 
condition of conservation of grains Eq. (\ref{e2001}) we obtain last relations 
immediately.

\subsection{Time dependent input flow}

As we have noted in Sec. I, space periodicity of the pile evolution may 
disappear if an input flow increases with time. Let us consider the simplest 
example of a linear time dependence: 
$r_0(t) = v_0 t$ (the constant $v_0$ has dimensionality of a velocity). Now 
there are not any jumps of profiles at the front point, since the growth of 
$r_0$ starts from zero and an initial state is empty, so the problem is simpler 
than that considered above. 

Instead of trying to solve the problem directly, one may proceed in the 
following way. First we suppose that the velocity of the front is constant. Let 
us call it $v_d$ again. From a moving boundary condition $h(x=-v_d t,t \geq 
0)=0$ at the front point accounting for Eq. (\ref{e2}) we obtain $w(z)$. Then 
from another moving boundary condition $r(x=-v_d t, t\geq 0)=0$ and 
Eq.(\ref{e2}) one may get $u(z)$. Inserting both answers into the condition 
$r(x=0,t) = v_0 t$ we (i) verify that the choice of a constant front 
velocity is right and (ii) find $v_d$ as a function of $v_0$.

We write out the answers immediately and check them by substituting in Eq. 
(\ref{e1}). 
The profiles of moving and static grains look as 

\begin{eqnarray}
\label{i1}
r & = & \frac{v_0}{v_d}\,(x + v_d t) \ ,
\nonumber \\
h & = & \frac{v_u}{v_d+v_u}\,\theta_f(x + v_d t) 
\end{eqnarray}
(see Fig. 4), where the front velocity is

\begin{equation}
\label{i2}
v_d = \frac{1}{2}\,\frac{v_u-v}{1+\theta_f v_u/v_0}\left[\sqrt{1+4\frac{v 
v_u}{(v_u-v)^2}\!\left(\! 1+\theta_f\frac{v_u}{v_0}\! \right)}-1\right]  . 
\end{equation}
In the figure, the evolution of the pile in the situation under consideration is 
depicted. One may see that the relation between amounts of rolling and static 
grains depends considerably on the rate of the input flow increase. As it 
follows from Eqs. (\ref{i1}) and (\ref{i2}), when 
$v_0 \gg v_u \theta_f$,
  
\begin{eqnarray}
\label{i3}
v_d & \cong & v \ ,
\nonumber \\
r   & \cong & \frac{v_0}{v}\,(x + v t) \ ,
\nonumber \\
h   & \cong & \frac{v_u}{v+v_u}\,\theta_f(x + v t) \ ,
\end{eqnarray}
and the amount of rolling grains is much higher than the amount of static grains 
[Fig. 4(a)]. The front velocity approaches its highest possible value, and 
rolling grains have no time to converse into static ones. For a low $c$ 
  
\begin{eqnarray}
\label{i4}
v_d & \cong & \sqrt{\frac{v_0 v}{\theta_f}} \ ,
\nonumber \\
r   & \cong & \sqrt{\frac{v_0 \theta_f}{v}}\, (x + \sqrt{\frac{v_0 v}{\theta_f}} 
t) \ ,
\nonumber \\
h   & \cong & \theta_f(x + \sqrt{\frac{v_0 v}{\theta_f}} t) \ ,
\end{eqnarray}
so the front velocity tends to zero, the slope of the distribution of static grains 
is nearly critical, and the relative amount of rolling grains is small [Fig. 
4(b)]. 

The amounts of rolling and static grains in a pile are equal for the following 
particular value of the rate of the input flow growth: 

\begin{equation}
\label{i5}
v_0^\ast = \frac{v v_u}{v+2v_u}\theta_f 
\end{equation}
(compare with Eq. (\ref{e5005}) from Sec. II\,A). In this particular case, the 
front velocity is $v_d^\ast=v/2$.

\section{Sandpile evolution starting from critical state}

Now, let us switch on the input flow of grains when there is 
already a pile with the critical slope. Thus, the initial condition for $h$ is 
$h(x,t=0)=\theta_f(x+d)$, where $d$ is the initial horizontal size of a pile.
All other initial and boundary conditions are the same as in Sec. II\,A (the 
input flow is time-independent). 

Now one can proceed with the calculations similar to those that were made in Sec. II 
to obtain the solutions describing the evolution of the pile. We omit the tedious 
calculations and present immediately results for the case under consideration. 
Main answers are presented schematically in Fig. 5 in which the distributions of 
rolling grains $r(x)$ and the static ones $h(x)$ are shown in several successive 
moments. (Note that scales change from one figure to another.) Trajectories of 
the front and the breaks on the $t,x$-plane are shown in Fig. 6.

Let us comment the contents of Fig. 5, since it looks rather intricate. At the 
initial moment rolling grains are absent, and the angle of the profile of static 
grains is critical. After we switch on an external flow, a step of rolling 
grains starts to descend downhill the critical profile of static grains [Fig. 
5(a)]. Rolling grains do not convert to the static ones. 

After the step approaches the last point of the profile of static grains [Fig. 
5(b)] it proceeds to move to the left with the same velocity $v$, but the height 
of the jump starts to decrease with time [Fig. 5(c)]. A new linear part of 
$r(x)$ emerges and begins to move uphill with the velocity $v_u$. (Recall that 
all linear parts of profiles move as a whole or stay without movement. They can 
not change their slope with time.) A new more gently sloping part of $h(x)$ 
simultaneously appears and begin to move to the left with the velocity $v$, so a 
break of $h(x)$, coinciding with the break of $r(x)$, moves uphill with the 
velocity $v_u$. 

After the decreasing jump at the front of $r(x)$ disappears [Fig. 5(d)] the 
linear part of $r(x,t)$ proceeds to move uphill with the velocity $v_u$ [Fig. 
5(e)]. The corresponding part of $h(x)$ also proceeds to move uphill. Behind it, 
a new static part of the profile with the critical slope emerges, so, in fact, 
an inclined step moves uphill.

When its first point touches a wall [Fig. 5(f)], a new static linear part of 
$r(x)$ appears to satisfy the boundary condition $r(x=0,t)=r_0$ [Fig. 5(g)]. An 
old part of $r(x)$ proceeds to move uphill, so a new break of $r(x)$, that moves 
downhill with the velocity $v$, emerges. Evolution of $h(x,t)$ proceeds as before. 
At some instant [Fig. 5(h)] the last point of $r(x)$ and the break of $r(x)$ meet. 
New linear part of $r(x)$ emerges and starts to move downhill [Fig. 5(i)]. The 
$h(x)$ profile evolves in the same way as before. 

When all the $h(x)$ profile appears to be critical [Fig. 5(j)], the whole $r(x)$ 
distribution is linear.  This inclined profile starts to move down the critical 
one [Fig. 5(k)]. The only difference with the instant depicted in Fig. 6(a) is 
that the step of rolling grains is inclined now. After its first point approaches 
the last left point of the critical pile [Fig. 5(l)] new linear parts of $r(x)$ 
and $h(x)$ emerge which move to the left with some velocity $v_d < v, v_u$ [Fig.  
5(m)]. Two breaks of $r(x)$ emerge which move one to each other. 

After the breaks meet [Fig. 5(n)], two new breaks of $r(x)$ emerge which move away 
one from each other [Fig. 5(o)] while $h(x,t)$ proceeds the previous evolution. 
The moving downhill break of $r(x)$ overtakes the front at some moment [Fig. 
5(p)]; after that an inclined step starts to climb uphill 
leaving after itself a new static critical slope part [Fig. 5(q)]. Afterwards 
the evolution proceeds in the same way.  

Thus, one sees that the evolution of a pile in this case looks more complicated 
than in Sec. II. Nevertheless, as it is evident from Fig. 6, in which all 
trajectories of the front and breaks of $r(x,t)$ and $h(x,t)$ are shown, a 
general similarity remains. 

Let us describe Fig. \ref{fig6}. The solid lines $x=v_u(t-t_{2m}^-), v_u(t-
t_{2m})$, $m=0,1,\ldots$ show trajectories of the $r(x,t)$ and $h(x,t)$ breaks 
moving uphill (we introduce these notations for the instants of time to keep a 
tie with the corresponding notations in Sec. II). The dashed lines $x=-v(t-
t_{2m}^-), -v(t-t_{2m})$, $m=0,1,\ldots$ and $x=-vt$ depict trajectories of the 
$r(x,t)$ break moving downhill. Comparing with Fig. 2 from the previous section 
one sees that the corresponding lines for the break trajectories are splitted 
now, and static segments of the front trajectory appear. The expressions for 
particular times shown in Fig. 6 are

\begin{eqnarray}
\label{e13001}
t_{2m-2}=m\frac{v+v_u}{v v_u}d +
\frac{(m-1)m}{2}\frac{(v+v_u)^2}{v v_u^2} , \nonumber \\[2pt]
t_{2m}^-=m\frac{v+v_u}{v v_u}d +
\frac{m(m+1)}{2}\frac{(v+v_u)^2}{v v_u^2} , \nonumber \\[2pt] 
m=1,2,\ldots
\end{eqnarray}
-- for the instants when the breaks of $r(x,t)$ and $h(x,t)$ reach a wall -- and

\begin{eqnarray}
\label{e13002}
t_{2m-1}  =  \frac{{\cal V}_{m}}{v v_u}d  +
\frac{m}{2}\frac{v+v_u}{v v_u^2} [(m-1)v + (m+1) v_u]
\frac{r_0}{\theta_f} , \nonumber \\[2pt]
t_{2m+1}^-  =  \frac{{\cal V}_{m}}{v v_u}d +
\frac{m+1}{2}\,\frac{v+v_u}{v v_u^2} [m v + (m+2)v_u]
\frac{r_0}{\theta_f} , \nonumber \\[2pt]  
m=0,1,\ldots
\end{eqnarray}
-- for the times at which the break of $r(x,t)$ overtakes the front.

The trajectory of the front can be described by the following relation 

\begin{eqnarray}
\label{e13003}
x(t_{2m+1}^-<t<t_{2m+1}) = -d - (m+1)\frac{v+v_u}{v_u}\frac{r_0}{\theta_f},  
\nonumber \\[2pt]
x(t_{2m-1}<t<t_{2m+1}^-) = 
\nonumber \\[2pt]
-\frac{m(m+1)}{2}\frac{(v+v_u)^2}{v_u{\cal V}_{m}}\frac{r_0}{\theta_f}- 
\frac{v v_u}{{\cal V}_{m}}t , \nonumber \\ 
m=0,1,\ldots
\end{eqnarray}
and $x(0<t<t_{-1})=-d$. Note that the second relation of Eq. (\ref{e13003}) 
coincides with Eq. (\ref{e6}). Thus, at the corresponding instants, the front 
velocities are the same in both cases. The first relation of Eq. (\ref{e13003}) 
again demonstrates the space periodicity of the process.

Time intervals for which the front is motionless do not change with time:

\begin{eqnarray}
\label{e13004}
t_{2m-1}-t_{2m-1}^- = 
\frac{v+v_u}{v v_u}d \ , \ m=1,2,\ldots 
\end{eqnarray}
These intervals appear because of a new initial condition,
so even at long times one may observe the influence of the initial condition. 
Nevertheless, the duration of the other time intervals, when the front moves, 
increases with time. Thus, the relative effect of the initial conditions 
decreases with time.   

We shall need only the following result for $h$:

\begin{eqnarray}
\label{e13005}
h(v_u(t-t_{2m+2}^-)<x<v_u(t-t_{2m+2})) = \nonumber \\[2pt] 
\theta_f \left( x+d+(m+1)\frac{v+v_u}{v_u}\frac{r_0}{\theta_f} \right),  
\nonumber \\[2pt]
h(v_u(t-t_{2m+2}^-)<x<v_u(t-t_{2m})) = \nonumber \\[2pt] 
\frac{m}{2}\frac{(v+v_u)}{v_u}r_0 +  
\frac{{\cal V}_{m}}{(m+1)(v+v_u)}\,\theta_f \left( x + \frac{v v_u}{{\cal 
V}_{m}}t \right) 
, \nonumber \\[2pt] 
m=0,1,\ldots
\end{eqnarray}
One sees that these relations are valid for different strips $v_u(t-t_{2m+2}^-
)<x<v_u(t-t_{2m+2})$ and $v_u(t-t_{2m+2}^-)<x<v_u(t-t_{2m})$ of the $x,t$-plane.
The first relation of Eq. (\ref{e13005}) describes parts of the distribution of 
static grains with the critical slope, and the second relation of Eq. 
(\ref{e13005}) coincides with the corresponding answer Eq. (\ref{e10}) for the 
first case. 

The slope angles of the profiles of static grains obtained from Eq. 
(\ref{e13005}) are shown in Fig. 7 which demonstrates most naturally the 
relaxation to the critical state. Unlike the previous section (see Fig. 2), in 
this case, segments with the critical angle are present. As can be seen 
from Fig. 7, different segments may overlap, and the profile may have two or even 
three parts with different slopes simultaneously (see also Fig. 6). Long time 
asymptotes of the pile slope are the same as in section II.

\section{Conclusions}

In summary, in the case of the thick flow regime, we have shown that space 
periodicity takes place during a sandpile evolution even for the one-component 
pile. The pile spills during a repeating process if an income flow is constant: 
grains are piling layer by layer. The thickness of the layers coincides 
surprisingly with the thickness of stratified layers at the two-component 
sandpile problem \cite{makse97prl,ciz}. Thus, in the one-component pile, we 
found a clear precursor of the stratification phenomena. 

We have found very reach behavior using the most simple and clear approach 
\cite{bout1} admiting an easy analytical treatment.
The slope of the pile goes to its final critical value after complex relaxation 
which is a long discontinuous process. Indeed, in the case of a constant income 
flow, the pile slope may change only discontinuously (see Figs. \ref{fig2} and 
\ref{fig6}). At long times the deviation of the slope from its critical value is 
proportional to $\sqrt{r_0/t}$. 

The pile evolution from different initial conditions looks very similar at long 
times but even at such times one can observe some difference (compare Figs. 
\ref{fig2} and \ref{fig6} for two most natural situations: the sandpile 
evolution starting from an empty state and the evolution starting from a 
critical state). 

The space periodicity disappears if an input flow grows linearly with time. In 
such a situation the sandpile front moves with a constant velocity all the time, and 
the pile evolves continuously without any peculiarities (see Fig. \ref{fig4}).

Formally speaking, we studied only the one-dimensional problem, but an ordinary 
three-dimensional sandpile with an axial symmetry (that means that sand is 
pouring on an infinite horizontal plane at a single point) may be described by 
the same equations as Eqs. (1) (the coordinate $x$ means the distance from the 
center). Thus, our results also stay valid for such a pile if one applies 
the phenomenological description \cite{bouch,bout1}. 

Aradian, Rapha\"{e}l, and de Gennes \cite{aradian} have recently 
introduced a dependence of 
the downhill velocity $v$ on $r$ in the frames of the thick flow regime 
($v_u=\mbox{const}$). The profile shapes in that case become nonlinear 
but a general picture of a sandpile evolution will hardly change, since the 
structure of the evolution equations is the same -- in fact, the second equation of 
Eqs. (\ref{e1}) is independent of the first one.

The following questions remain open. How crucially do our results depend on the 
used, maybe oversimplified, phenomenological approach? Are they really of 
general significance?
Does the sandpile evolution in a thin flow regime differ essentially from the 
described one? 
It is apparently impossible to answer these questions within the present 
approach, however the last results \cite{mahad} and \cite{emig} 
on granular flows in thin and intermediate regimes let one hope to solve these 
beautiful problems. \\

\section*{Acknowledgments}

SND thanks PRAXIS XXI (Portugal) for a research grant PRAXIS XXI/BCC/16418/98. 
JFFM was partially supported by the projects PRAXIS/2/2.1/FIS/299/94, 
PRAXIS/2/2.1/FIS/302/94 and NATO grant No. CRG-970332. We also thank M.C. Marques 
for reading the manuscript and A.V. Goltsev,  Yu.G. Pogorelov, and A.N. Samukhin 
for many useful discussions. \\
$^{\ast}$      Electronic address: sdorogov@fc.up.pt\\
$^{\dagger}$   Electronic address: jfmendes@fc.up.pt

\end{multicols}

\end{document}